\documentclass[preprint]{aastex}

\shorttitle{Spectroscopic survey of NGC~6871}
\shortauthors{Balog et al.}

\begin{document}

\title{Spectroscopic Survey of the Galactic Open Cluster NGC~6871 I. New Emission Line Stars}

\author{Zoltan Balog\altaffilmark{1} and Scott J. Kenyon}
\affil{Smithsonian Astrophysical Observatory, 60 Garden Street, Cambridge, MA 02138}
\email{zbalog@cfa.harvard.edu and kenyon@cfa.harvard.edu}

\altaffiltext{1}{on leave from Dept. of Optics and Quantum Electronics University of Szeged, Szeged, Hungary}

\begin{abstract}
We analyze spectra of 44 emission line stars detected in a low resolution optical spectroscopic survey of the galactic open cluster NGC~6871. The survey of 1217 stars is complete to $V = 14.9$ and includes stars with $V < 16.5$ between the zero age main sequence (ZAMS) and the $10^7$ yr pre-main sequence (PMS) isochrone. Of the 44 emission line stars in this survey, 28 show obvious emission in $H\alpha$ and 16 have weak $H\alpha$ absorption (compared to $H\gamma$). We use the reddening to separate foreground and background stars from the cluster members; the position in the HR diagram (HRD)  or the presence of forbidden emission lines ([NII], [SII]) then yields the evolutionary status of the emission line stars. A comparison of the $H\alpha$ spectral index distribution in NGC~6871 with the distribution of young stars in the Taurus-Auriga molecular cloud indicates that the late type PMS stars in our sample may be weak line T Tauri stars (wTTs). Many of these stars show [SII] ($36\%$) and [NII] ($45\%$) emission.
\end{abstract}


\section{Introduction}
Establishing mass-and age-dependent color-luminosity relations is important for understanding the physical processes leading from molecular clouds to main sequence stars. As in main sequence and post main sequence stars, observational HR-diagrams for PMS stars allow comparisons between observations and predictions of evolutionary models. Because isochrone fitting allows a more reliable measurement of distances and ages for star clusters than for single stars, the detection of PMS stars as members of star clusters offers an advantage over the observation of individual stars in more extended star forming regions. This advantage leads to accurate colors and luminosities for the possible PMS members of open clusters with well defined constraints on age and mass. 

The aim of this paper is to describe 37 new emission line objects along the line of sight to the galactic open cluster NGC~6871. We identify these stars from a program to measure accurate spectral types, to derive the reddening, to separate cluster members from foreground and background stars, and to construct the Hertzsprung-Russell diagram (HRD) for cluster members. We will describe these results in a future paper.

With an age of roughly $10^7$ yr, NGC~6871 is at an important stage in the early evolution of an open cluster. Younger clusters with ages of $\sim 10^6$ yr are often embedded in a bright nebulosity or a dark molecular cloud. Most of the young stars in these clusters have circumstellar disks \citep{hais01}. Older star clusters with ages of $10^8$ yr have little nebulosity or molecular gas. Few stars in these cluster have disks \citep{habi01,span01}. In NGC~6871, the lack of significant nebulosity or molecular gas allows optical spectroscopic observations to derive spectral types and luminosities of cluster members, and to measure IR excesses from circumstellar disks. Optical data provide a reliable measurement of the IMF for comparison with other young clusters \citep[e.g.,][]{sles02}. Recent work suggests that circumstellar disks disappear on timescales of 5-10 Myr \citep{hais01}. Observations of IR excess emission in NGC6871 thus yield better constraints on disk timescale.

NGC~6871 has been relatively well studied photometrically. The cluster is part of the Cyg OB3 association \citep{garm92}, one of the largest nearby star formation complexes. The most recent UBV CCD photometry published by \citet{mass95} has a limiting magnitude $V \simeq16.5$. Several age estimates exist for the cluster members. From isochrone fitting, \citet{mass95} derived ages of 2 - 5 Myr for the stars with $M > 25 M_\odot$ (spectral type B0 and earlier). From uvby photometry for stars with spectral type earlier than B5 ($M > 4M_\odot$), \citet{reim89} estimated an age of 10 - 20 Myr. The difference between the two ages suggests either a large age spread among the cluster members, contamination from nearby stars in the local spiral arm \citep{sles02}, or large uncertainties in age estimates derived from optical photometry. The presence of O stars in the cluster sets an upper limit for ages of high mass stars. According to \citet{mass95}, the highest mass stars in the cluster have $M = 40~M_{\odot}$ and main sequence lifetimes of 4-5 Myr \citep{schal92}. However, \citet{mass95} note that the cluster contains evolved 15 $M_\odot$ stars with main sequence lifetimes of 11 Myr. 

One of the goals of our program is to test the origin of the age difference by deriving accurate spectral types for cluster members. According to the WEBDA catalog of J-C. Mermilliod \citep{merm95}, 12 stars in the cluster have reliable spectral types \citep[e.g.,][]{rom51, hilt56,wal71,mass95}. The SIMBAD database lists another 151 stars with spectral classifications in the field of NGC~6871. Only 58 of these stars are included in the photometric database of \citet{mass95} from which we selected our spectroscopic candidates. The discrepancy between the numbers of stars with spectral types in different catalogues can be explained by the incompletness of the WEBDA and the uncertain spatial bounderies of the catalogue of \citet{mass95}. Some stars with spectral type must lie outside the CCD field of \citet{mass95}. Our sample of 1217 stars will yield a good comparison between ages derived from photometric and spectroscopic techniques. This large sample also provides a good measurement of the frequency of $H\alpha$ and other emission lines as a function of spectral types. Emission lines disappear when disk accretion ends; our data thus provide an estimate of the time scale for the end of disk accretion for intermediate and low mass stars. 

We adopt an automatic approach and measure spectral types from spectral indices derived from digital spectra. This method allows robust measurement of spectral types and can be rapidly applied to large numbers of spectra. The accuracy of the method is comparable to visual estimates. We also use spectral indices to identify emission line objects. We compare the strength of the H$\alpha$ and H$\gamma$ indices and identify an object as emission line star if the H$\alpha$ absorption is weak relative to H$\gamma$. This method yields an automatic identification of emission line stars in cases where the emission is not obviously detectable by visual examination. 

In this first paper of the series we report the discovery of 37 new emission line stars and analyze their spectra. Eight stars with accurate spectral types are known emission line stars (we detect seven of these stars). Here, we extend this sample to 45 stars and begin to separate cluster members from foreground and background stars.

\section{Observations}

We use the WEBDA for selecting the target stars for our survey and adopt the WEBDA designation for stars throughout this paper. We selected stars from the WEBDA which have reliable UBV CCD photometry. The survey of 1217 stars is complete to $V = 14.9$ and includes stars with $V < 16.5$ between the ZAMS and the $10^7$ yr PMS isochrone of \citet{siess00}.

P. Berlind, M. Calkins, and several other observers acquired low resolution optical spectra of stars on the field of NGC~6871 with FAST, a high throughput, slit spectrograph mounted at the Fred L. Whipple Observatory 1.5-m telescope on Mount Hopkins, Arizona \citep{fab98}.
They used a 300 g mm$^{-1}$ grating  blazed at 4750 \AA, a 3\arcsec~slit, and a thinned Loral 512 $\times$ 2688 CCD.  These spectra cover 3800--7500 \AA~at a resolution of $\sim$ 6 \AA.  We wavelength-calibrate the spectra in NOAO IRAF\footnote{IRAF is distributed by the National Optical Astronomy Observatory, which is operated by the Association of Universities for Research in Astronomy, Inc. under contract to the National Science Foundation.}.
After trimming the CCD frames at each end of the slit, we correct for the bias level, flat-field each frame, apply an illumination correction, and derive a full wavelength solution from calibration lamps acquired immediately after each exposure.  The wavelength solution for each frame has a probable error of $\pm$0.5--1.0 \AA.  To construct final 1-D spectra, we extract object and sky spectra using the optimal extraction algorithm within APEXTRACT.  Most spectra have moderate signal-to-noise, S/N $\gtrsim$ 30 per pixel.

To measure the spectral types of individual stars, we measure continuum magnitudes and indices using narrow passbands \citep{ocnl73,worth94}. 
The absorption and emission indices, derived using SBANDS within IRAF, are $I_{\lambda}$ = $-$2.5 log($F_{\lambda}$/$\bar{F}$), where $F_{\lambda}$ is the average flux in the passband $\lambda$ and $\bar{F}$ is the continuum flux interpolated between the fluxes in the neighboring blue band centered at $\lambda_b$ and the red band centered at $\lambda_r$. Table 1 lists the indices used in our analysis. We establish relations between indices and spectral type using the standards of \citet{jac84}. We use the sum of the He indices ($I_{He}=I_{He\lambda4144}+I_{He\lambda4387}+I_{He\lambda4471}+I_{He\lambda4922}$) for stars earlier than B2, and two different sums of the H indices ($I_{H1}=I_{H\alpha}+I_{H\beta}+I{H\gamma}+I_{H\delta}+I_{H\lambda3889}+I_{H\lambda3835}$, $I_{H2}=I_{H\alpha}+I_{H\beta}+I_{H\gamma}+I_{H\delta}+I_{H\lambda3889}$) for stars later than B0. We define two H indices because $H\lambda3835$ is unreliable for stars with spectral type of A2 or later. We use the sum of TiO indices ($I_{TiO}=I_{TiO\lambda5968}+I_{TiO\lambda6182}+I_{TiO\lambda6234}+I_{TiO\lambda7100}$) for stars with detectable TiO bands. We apply three different linear fits to the different parts of the index - ST relation for the hydrogen and the helium indices and a 3rd degree polynomial fit to the TiO index, where ST is the numerical spectral type ($\rm 0=O0, 10=B0, 20=A0$ and so on). 

To check the reliability of our spectral types, we compare with published types for 58 stars in our sample. We define $\delta$ST = ST(present paper) - ST(literature), and derive the mean $<\delta \rm{ST} >$ = $-0.21 \pm 4.1$. Twelve stars are responsible for most of the dispersion. Checking these types visually, we verify that all of these stars are misclassified. For example, of the three K5 stars in the SIMBAD database, one is an M star with strong TiO bands and two are obvious G-type stars. Removing these misclassifications from the complete sample, we measure $<\delta \rm{ST} >$ = $0.15 \pm 1.7$. An error of 1-2 subclasses is typical of visual classifications \citep[e.g.,][]{cuti99}.

Spectral indices provide an efficient method for deriving spectral types from low resolution digital spectra. As multi-object spectrographs become available on many telescopes, robust digital spectral types are necessary to place cluster HR diagrams on a reliable, repeatable scale. Our technique yields accurate spectral types for stars without emission lines. Because luminosity classification requires higher spectral resolution, this technique does not yield reliable luminosity classes. However, spectral indices are easily generalized to derive luminosity classes on higher resolution spectra. To derive spectral types for emission line stars, we follow \citet{jasch87} and estimate spectral type visually. Based on comparisons with published spectral types, our classification for these stars have an uncertainty of $\pm$ 0.9 subclasses

\section{Results}
\subsection{Selection of emission line stars}

\begin{figure}
\begin{center}
\plotone{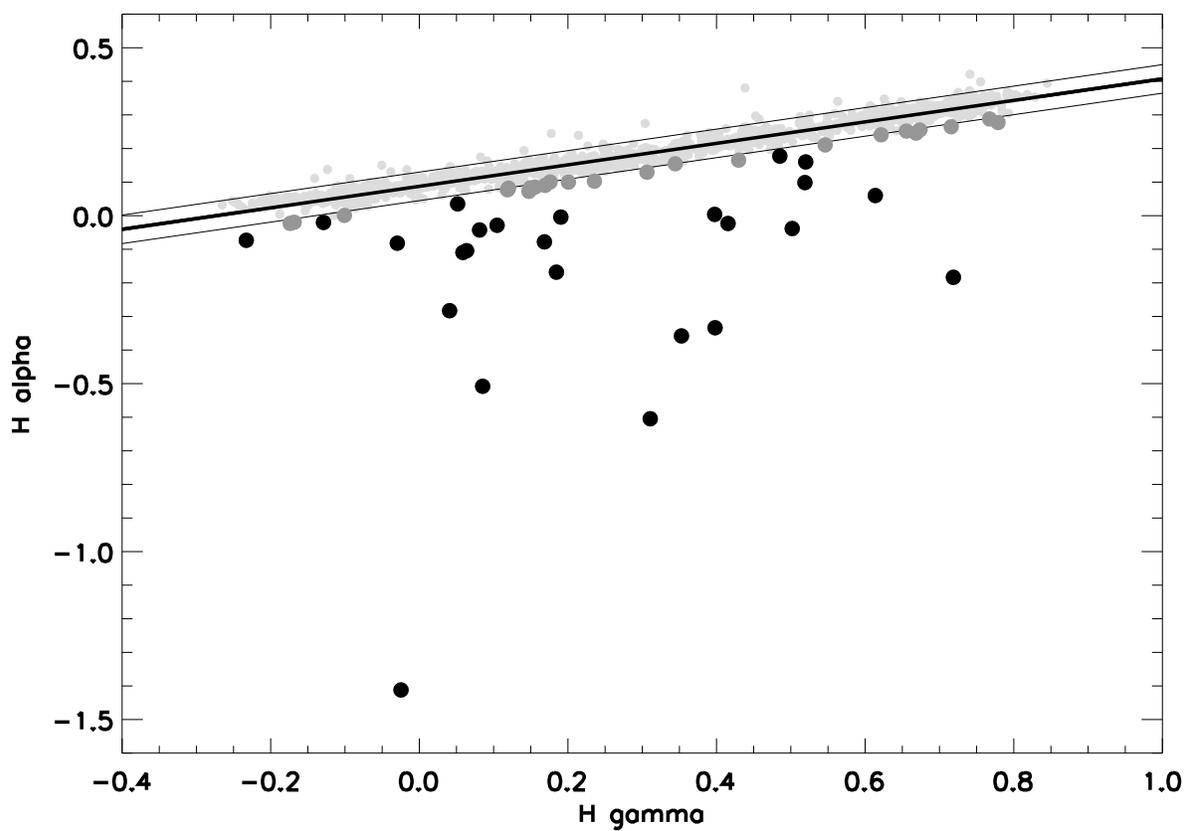}
\caption{$I_{H\alpha}$ vs $I_{H\gamma}$ diagram for stars in the spectroscopic sample. The lines represent the linear relation fitted to the indices (dark line) and the $2\sigma$ limits (light lines). Light grey dots denote the normal main sequence stars, grey dots denote the candidate emission line objects, and black dots denote the obvious emission line objects} 
\label{fig:1}
\end{center}
\end{figure}

We use HI indices to identify emission line stars in our sample. In normal main sequence stars, the HI equivalent widths are a function of spectral type \citep[see e.g.,][]{jasch87}. The equivalent widths peak at A2 spectral type, (e.g. $I_{H_\beta} \simeq 0.7$), and decrease towards earlier and later spectral types. All of the HI Balmer lines have roughly equivalent absorption line equivalent widths. In contrast, the equivalent widths of Balmer emission lines depend on the upper level. The relative fluxes for the H$\alpha$, H$\beta$, and H$\gamma$ emission lines are roughly 3:1:0.45 \citep{ost89}. Thus, HI emission fills in the H$\alpha$ absorption line before affecting higher members of the Balmer series. We use weak H$\alpha$ absorption relative to H$\beta$ and H$\gamma$ absorption to identify emission line stars with no obvious $H\alpha$ emission feature (see also \citet{bragg02}).

Fig 1. shows the $H\alpha$ index ($I_{H\alpha}$) as a function of the $H\gamma$ index ($I_{H\gamma}$) for our program stars. The majority of the stars follow a straight line with relatively small scatter. These are normal main sequence stars without emission lines. We derive a linear relation between the two indices for these stars with a least squares fit, 

$$I_{H\gamma}=0.32*I_{H\alpha}+0.09.$$

\noindent The rms of the fit is $\sigma = 0.02$. We make several iterations before accepting the final fit. First we reject the stars with obvious emission lines (negative $H\alpha$ indices). After performing a first fit we eliminate the stars which lie beyond the 2$\sigma$ limit. This group contains all of the obvious emission line stars and outliers with poor S/N spectra (above the 2$\sigma$ limit on Fig 1.). Then we perform a final fit and measure the rms of the linear relation for normal stars. We adopt this rms for selecting emission line candidates.  

We identify ``definite'' emission line objects as stars that fall more than 3$\sigma$ below the straight line fit. Candidate emission line objects fall 2-3$\sigma$ below the fit. Visual inspection of the spectra confirms that 22/24 of the definite emission line objects and 4/20 of the candidates have obvious $H\alpha$ emission. In the other stars, the $H\alpha$ absorption line is definitely weaker than in normal main sequence stars. Tables 2 and 3 list our sample of 44 emission line stars along with our derived spectral types, reddening, and  photometric data from the catalogue of \citet{mass95} (MJD95). We also list the designation of the stars from MJD95. Fig. 3 shows the color-magnitude diagram for these stars. 

\subsection{Analysis of the emission line stars}
In this section we derive the reddening $E(B-V)$ of stars in the field of NGC~6871 using their linked photometric and spectroscopic data. Accurate reddening values allow us to filter out foreground and background objects from our sample, and provide a more accurate placement of stars on the HRD.  

\subsubsection{Reddening and membership}

We calculate the $E(B-V)$ for each non-emission line (normal) main sequence star in our sample from the measured spectral type and the color table of \citet{keny95}. Because \citet{keny95} does not include stars with spectral type earlier than B0, we calculate a lower limit to $E(B-V)$ for these stars and do not include them in the analysis. The left panel of Fig. 2 shows the reddening histogram for normal main sequence stars in the cluster field.  
The distribution peaks around $E(B-V)=0.4$. This value agrees with previous values of $E(B-V)=0.443$ (WEBDA) and $E(B-V)=0.46$ \citep{mass95}. 

We use the reddening calculated by the computer code of \citet{hak97} to check our results. The \citet{hak97} program yields visual extinction as a function of Galactic longitude, Galactic latitude, and distance. The code combines the results of  published studies (see references in \citet{hak97}). Three studies \citep{fitz68,neck80,arn92} provide information on the reddening for the Galactic coordinates of NGC~6871. Our estimate for $E(B-V)$ agrees only with the results of \citet{neck80}. The study of \citet{arn92} overestimates the reddening by a factor of 2; the study of \citet{fitz68} underestimates the reddening by a factor of 2. We use the results of \citet{neck80} below.   

The right panel of Fig. 2 shows the reddening histogram of emission line stars in the cluster field. The distribution is similar to the histogram for normal stars. However there is an excess of emission line stars with virtually no reddening. These stars may be single young stars in the vicinity of the Sun or part of a nearby young association. Including these zero reddening stars, $22\%$ of the emission line stars seem to belong to the foreground population. The stars with $E(B-V)$ between 0.3 and 0.5 are probably members of the cluster. Roughly 1/3 ($38\%$) of our emission line stars belong to this group. Stars with $E(B-V)>0.5$ could be background objects. Roughly $40\%$ of the emission line stars has larger reddening than the cluster members. According to the $A_v$-distance relation of \citet{neck80}, these stars have distances larger than 1.9 kpc. The distance can be as large as 6.6 kpc for the highly reddened O star id2905 (HD226868). Such large distances seem improbable, because they imply unreasonably large absolute magnitudes. The calculated $z$ value above the galactic plane ($ z > 200$ pc) for id2905 also contrasts with the accepted $z$ values for O - B stars ($50 - 100$ pc), which suggests that this star has extra circumstellar reddening compared to cluster stars. 
The distances of the other background stars seem reasonable for the position of NGC~6871 but some of these stars might be cluster members with extra reddening due to circumstellar material. We examine this possibility further in the next subsection. 

\begin{figure}
\begin{center}
\plottwo{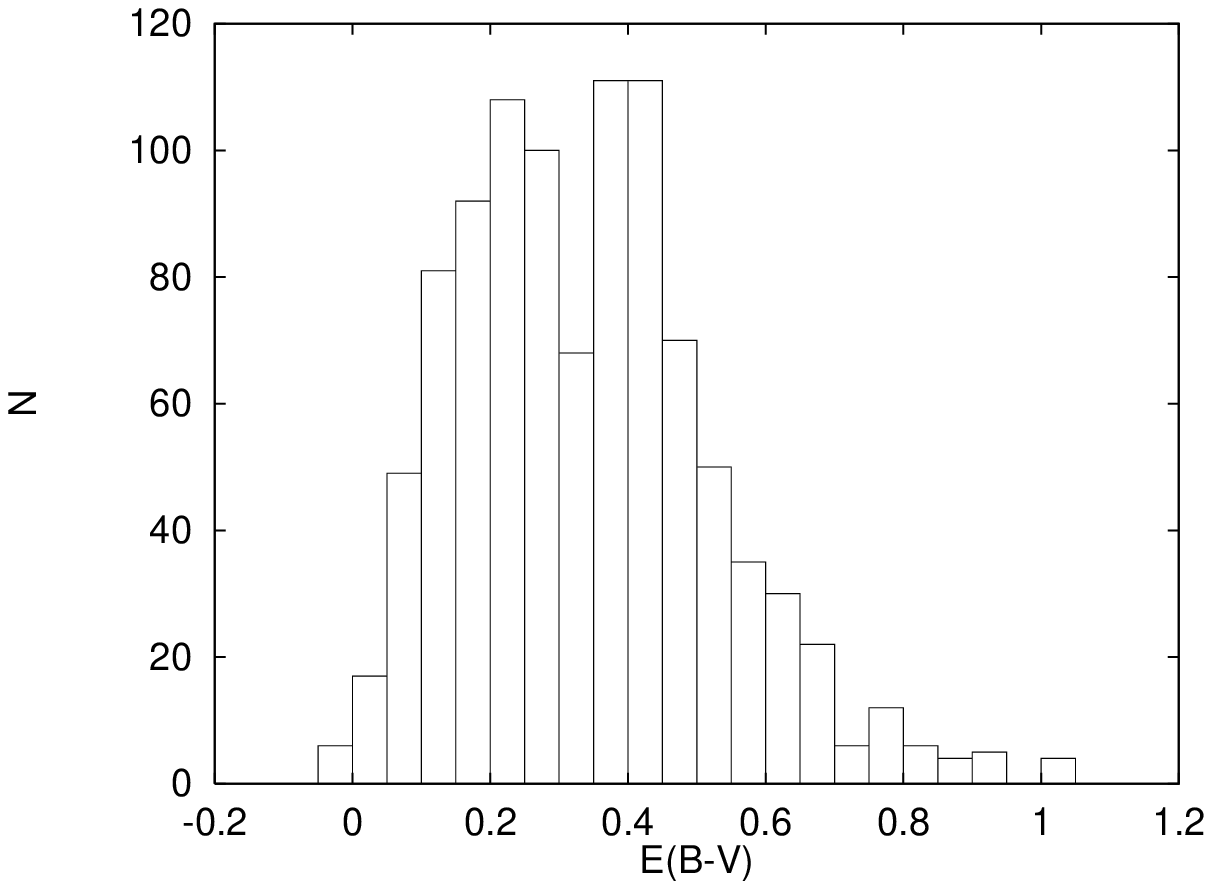}{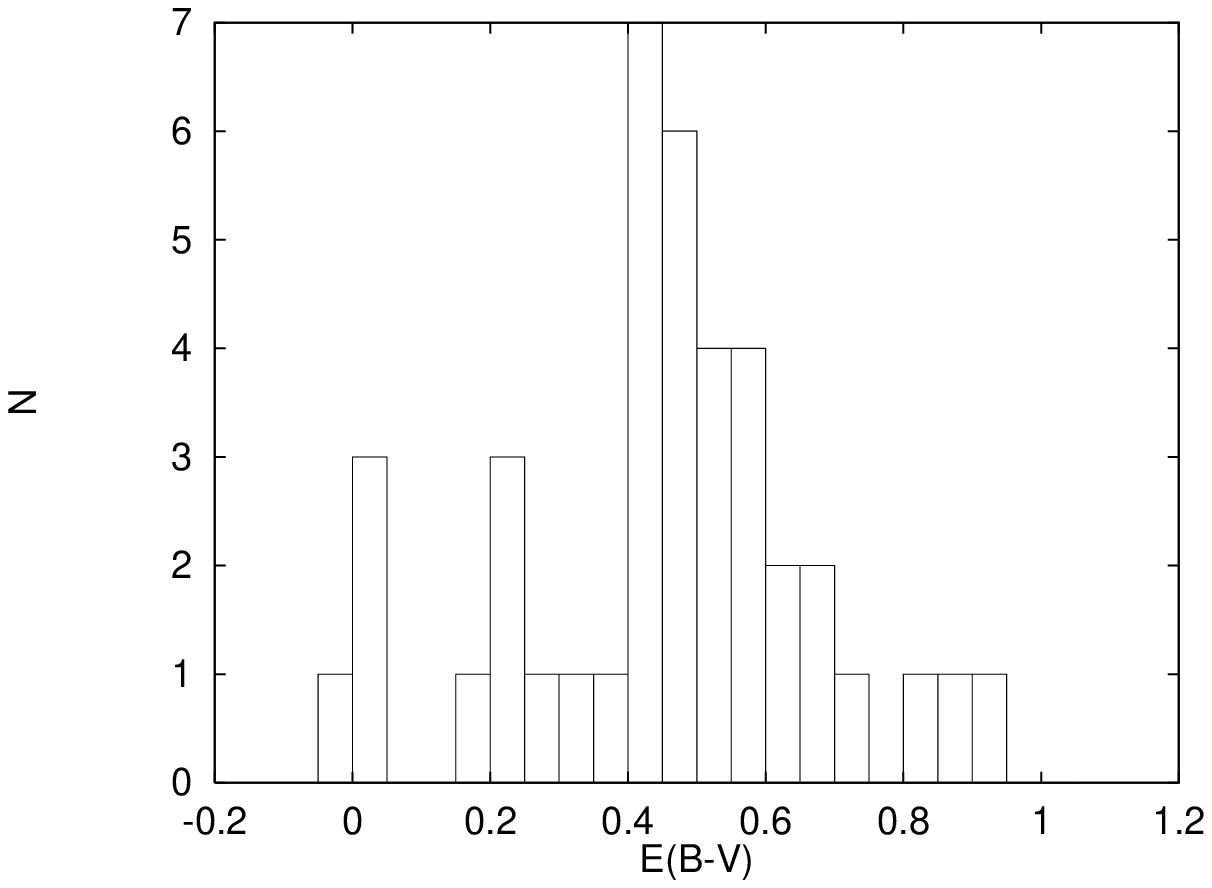}
\caption{Histogram of the reddening for normal main sequence stars (left panel) and emission line stars (right panel) on the field of NGC6871}
\label{fig:2}
\end{center}
\end{figure}

\subsubsection{Emission line objects on the cluster Hertzsprung-Russell diagram}

\begin{figure}
\begin{center}
\plottwo{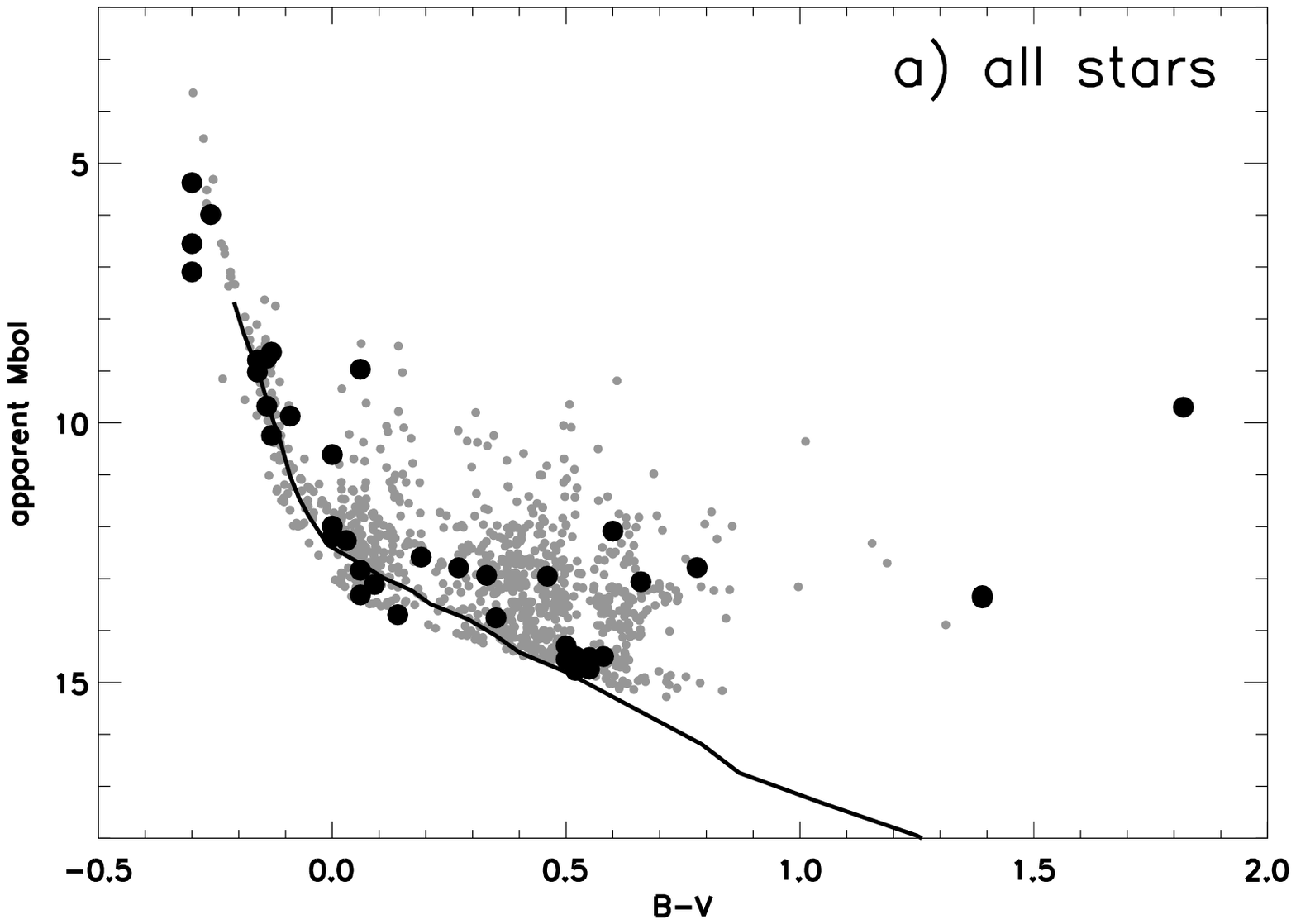}{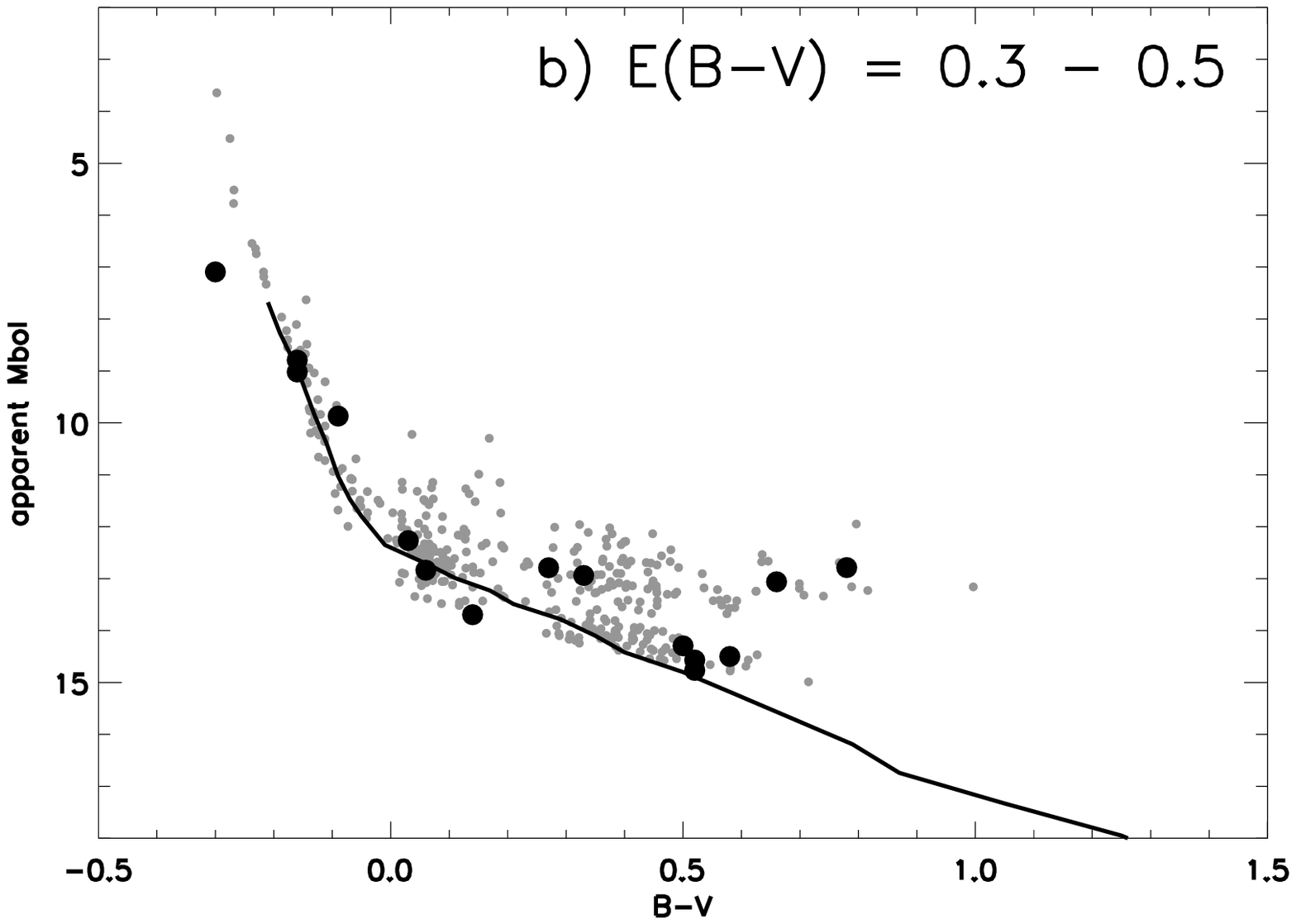}
\epsscale{2.2}
\plottwo{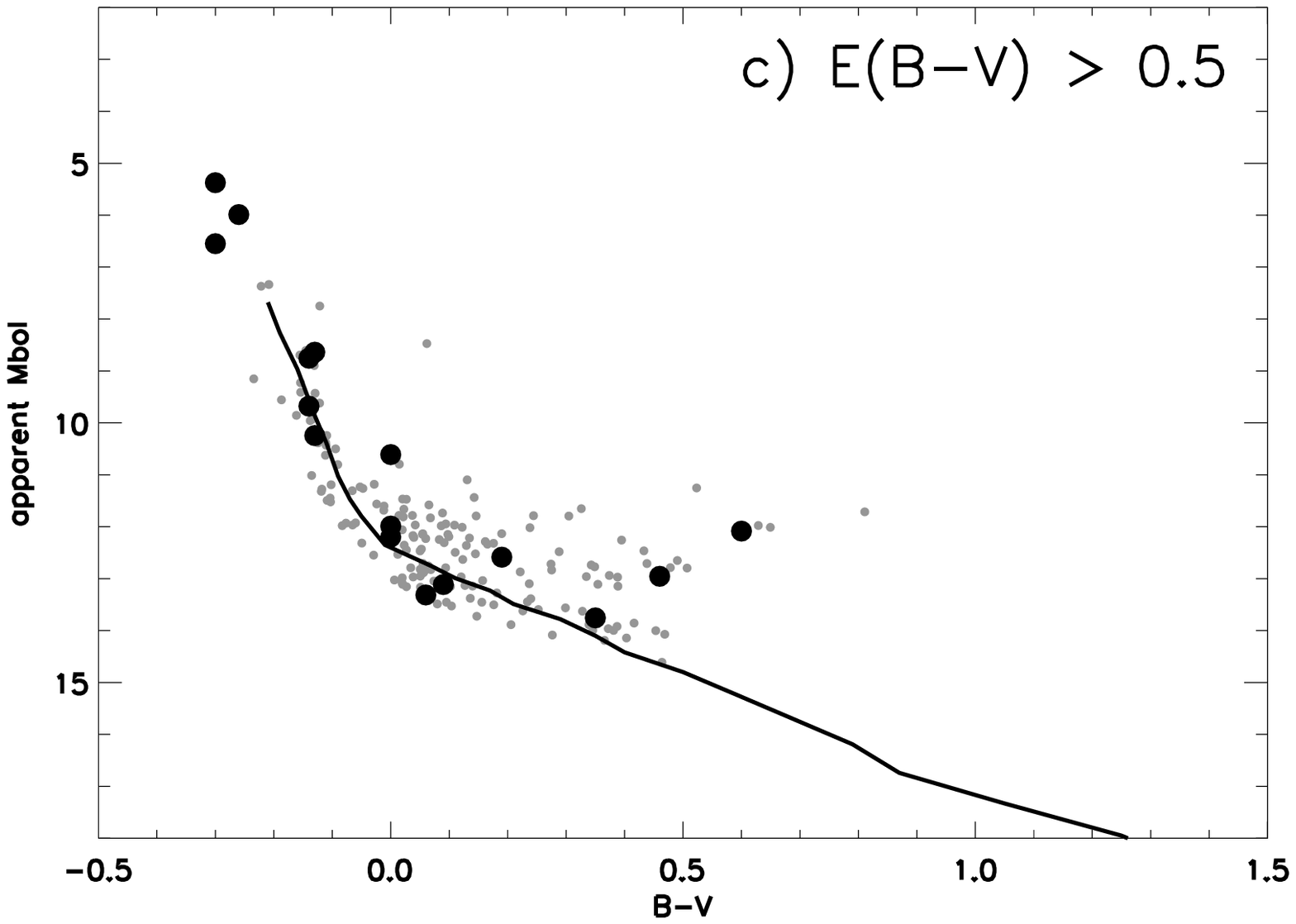}{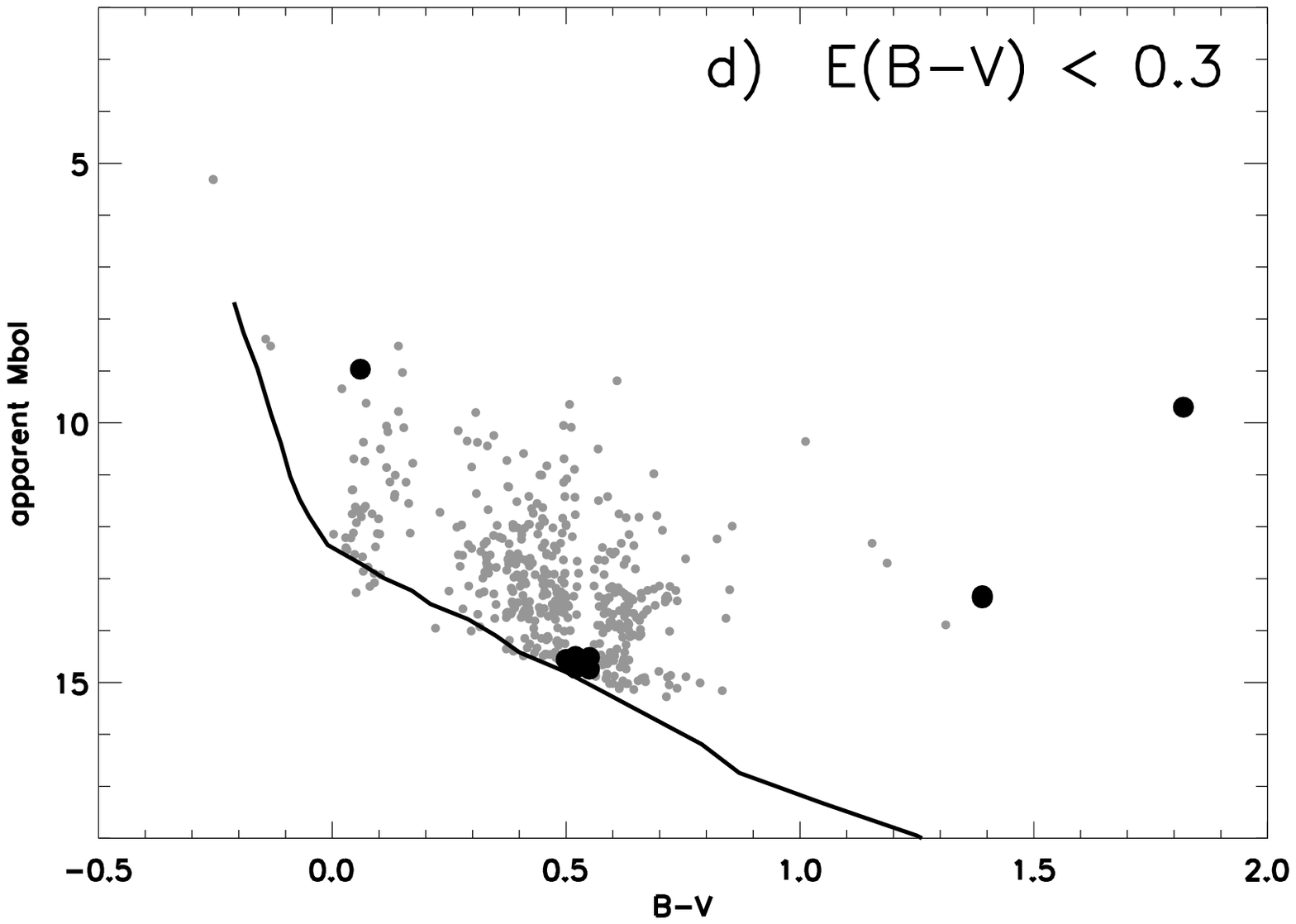}
\caption{HRDs of different subsamples of NGC~6871 based on the photometry of \citet{mass95}. Top left: all stars, top right: stars with $E(B-V)$ between 0.3 and 0.5, bottom left: stars with $E(B-V)>0.5$, bottom right: stars with $E(B-V)<0.3$. Small grey dots: normal main sequence stars, large black dots: emission line stars. The line represents the ZAMS of \citet{siess00} shifted with DM=11.0}
\label{fig:3}
\end{center}
\end{figure}

Having derived the B-V color excess for the emission line stars, we calculate the $A_v$ and the bolometric correction for each star in our sample and construct the HRD for each subsample specified in the previous section (foreground, cluster, background). Fig 3 shows these diagrams. Fig 3a (top left) shows the HRD for all stars in our sample, Fig 3b (top right) shows the HRD for the cluster; Fig 3c and Fig 3d (bottom left and right) show the HRD of the background and the foreground stars respectively. Small grey dots denote normal main sequence stars; large black dots denote emission line stars. The solid curves show the ZAMS of \citet{siess00} shifted with the distance modulus $DM=11.08$ which corresponds to a distance of 1649 pc \citep{bat91}. 

Although the majority of emission line stars in the cluster follow the ZAMS, four emission line stars lie above the main sequence. These objects may be PMS stars (see section 4.). Several other non-emission line objects lie close to these stars in the HRD. The relatively large scatter of these apparent PMS stars in the HRD suggest a large age-spread in the cluster \citep{mass95,reim89}. 

The HRD of background objects (Fig 3c) is very similar to the HRD for cluster members. Roughly half of the emission line stars lie on or above the ZAMS, which suggests that they are cluster members with extra reddening due to circumstellar material. Radial velocity measurements would test this possibility.

The stars with lower reddening occupy an area above the main sequence. Most of these stars are foreground objects with different distance moduli. Three of the four zero reddening stars may lie at the same distance: they are on a line nearly parallel to the ZAMS and may be a small group of young stars. The fourth of the unreddened stars may lie within 100 pc but its spectral type is uncertain. Further observations would place better limits on its distance. The other foreground stars lie just above the main sequence. Their position is consistent with the distance calculated from their reddening using the calibration of \citet{neck80} (1.2 kpc). They may be part of another young open cluster along the line of sight to NGC~6871 or a lower density concentration of stars in the local spiral arm.

\section{Short description of the spectra of the emission line stars}

In this section we give a short description of emission line stars in our sample. We follow the previous section and divide the sample into foreground, background, and cluster stars. In each section, we classify the stars as main sequence or pre-main sequence based on their positions in the HRD, the presence of [NII], [SII] emission lines, and the detection of the Li I $\lambda$6708 absorption line. Although Li~I~$\lambda$6708 absorption provides the best age measure for a young star, we detect Li I $\lambda$6708 only in a few cases due to the low resolution of our spectra. Even where we detect the Li I $\lambda$6708 line, the detection is very uncertain. Higher resolution spectra would provide a better test for the presence of Li I $\lambda$6708 in these stars.

With several O-type stars in the cluster, [NII] and [SII] emission lines could result from poor sky/nebula subtraction. To test this possibility, we measure an [SII] index for the complete spectroscopic sample. If fluctuations in nebular emission across the cluster field produce [NII] and [SII] emission, we expect roughly equal numbers of sources with apparent [NII] and [SII] absorption. Our test indicates a small dispersion in $I_{[SII]}$ with $\sigma = 0.01$. We identify a slight trend of increasing [SII] index towards later spectral types. The only exceptions to this trend are [SII] emission line stars. There are no stars with apparent [SII] absorption. Thus, we conclude that nebular contamination is not responsible for the [SII] or [NII] emission line stars. 

\subsection{Foreground stars}
There are 9 stars in our sample with low reddening. Because all are fainter than the Hipparcos limit, we cannot use accurate distances or proper motions to test the evolutionary status of these stars. Most have late spectral types (F - M). One (id2399) has an early spectral type (A2) with very weak  emission in the core of the $H\alpha$ line.

There are five F type stars among the foreground objects. Two (id1755, id1552) have visible emission; the other three (id1221, id2545, id1869) show weak $H\alpha$ absorption. These stars have roughly the same distance according to their position on the HRD and the $A_v$-distance relation of \citet{neck80}. This distance coincides with the distance of the majority of the normal foreground stars, suggesting that they may be part of a cluster between the Sun and NGC~6871. Radial velocity measurements would test this idea. Only one of these stars (id1755) is among our PMS candidates.

Two of the remaining three stars  (id1647, id1659) might be wTTs with spectral type around K8. All of their Balmer lines are in emission along with the Ca H and K lines. We detect Li I $\lambda$6708 line in the spectrum of id1659 even at our low resolution.

There is one M star in our sample (id113) with weak $H\alpha$ and [NII] emission. To examine the possibility that id113 is a dMe star, we estimate the number of nearby dMe stars from the observed stellar density of 0.084 star/$\rm pc^{-3}$ \citep{henr02}. Roughly $70\%$ of nearby stars are M dwarfs \citep{henr02}. At our completeness limit ($V \simeq 14.9$), we can detect M dwarfs at distances of 100 pc. The number of detectable M dwarfs towards NGC~6871 is two. The number of dMe stars among M dwarfs increases monotonically with spectral type ($4\%$ around M0 and $90\%$ M5 and later \citep{joy74}). The overall ratio of dMe stars among M dwarfs is $17\%$, which yields a modest probability ($\simeq 34\%$) of detecting one in our spectroscopic sample. Thus, id113 could be a dMe star. High resolution optical spectra of Li I $\lambda$6708 would discriminate between a field dMe star and a pre-main sequence star.

\subsection{Probable cluster members}
There are 15 emission line stars in this group. Their spectral types range from early B to late G. Most are definite emission line objects. More than half of the 15 cluster members (9 altogether) appear to be PMS stars. For 8 stars, we base this identification on their position on the HRD. One star has [NII] and [SII] emission.

There are three B type stars (id8, id186, id38) among the 11 definite emission line stars. They show emission only in $H\alpha$. Two of them (id8 and id38) were previously known \citep[see][]{grig88,bern01}. They could be classical Be stars.

There are two A stars (id2139, id1694) in this subgroup. One (id2139) shows signs of [NII] and [SII] emission. The $H\alpha$ emission of id1694 is very weak, but its position on the HRD indicates that it may be a PMS star.

There are three F type stars with definite emission features among the cluster members. Two (id2646, id2631) are PMS candidates with [NII] and [SII] emission. The third one (id2055) shows emission in $H\alpha$ and lies close to the main sequence. 

Three G type members (id2621, id2138, id2675) may be PMS stars. All lie well above the main sequence. The spectrum of id2621 contains [NII] and [SII] emission; we detect Li I $\lambda$6708 line in the spectrum of id2138. The third star of this group does not show any particular feature beside the weak emission in the core of the $H\alpha$ line.

\subsection{Stars with high reddening}

Based on their reddening, we identify 16 emission line stars as possible background objects. The majority are early type objects; only three have spectral type F or later. Roughly half of these stars lie above the ZAMS on the HRD, suggesting that these are probably cluster stars with extra (circumstellar) reddening.

Seven of the stars (id3, id6, id2352, id215, id2209, id1062, id1238) have B spectral types. All have definite emission lines except id3. Two (id1062 and id2209) lie above the main sequence and are included in our PMS candidate sample. In the case of id6 and id1062, the whole Balmer series is in emission; id2352, id215, id2209 and id1238 show detectable emission only in the $H\alpha$ line.

There are six A type stars in this group (id2320, id2492, id1651,id1093, id2637, id2156). Only two (id2320 and id2492) show definite $H\alpha$ emission; four have a weak $H\alpha$ absorption line. We identify four (id2320, id2492, id1651, id2156) as candidate PMS stars according to their position on the HRD. One of these stars (id2320) shows emission in [NII] and [SII] 

Because they lie above the main sequence, all of the three late type objects (id1379, id1845, id2245) are among our PMS candidates. Only one (id2245) shows Li I $\lambda$6708 absorption line.

\section{Comparing PMS candidates with the young stars in the Taurus-Auriga molecular clouds}

\begin{figure}[h]
\begin{center}
\epsscale{1.0}
\plottwo{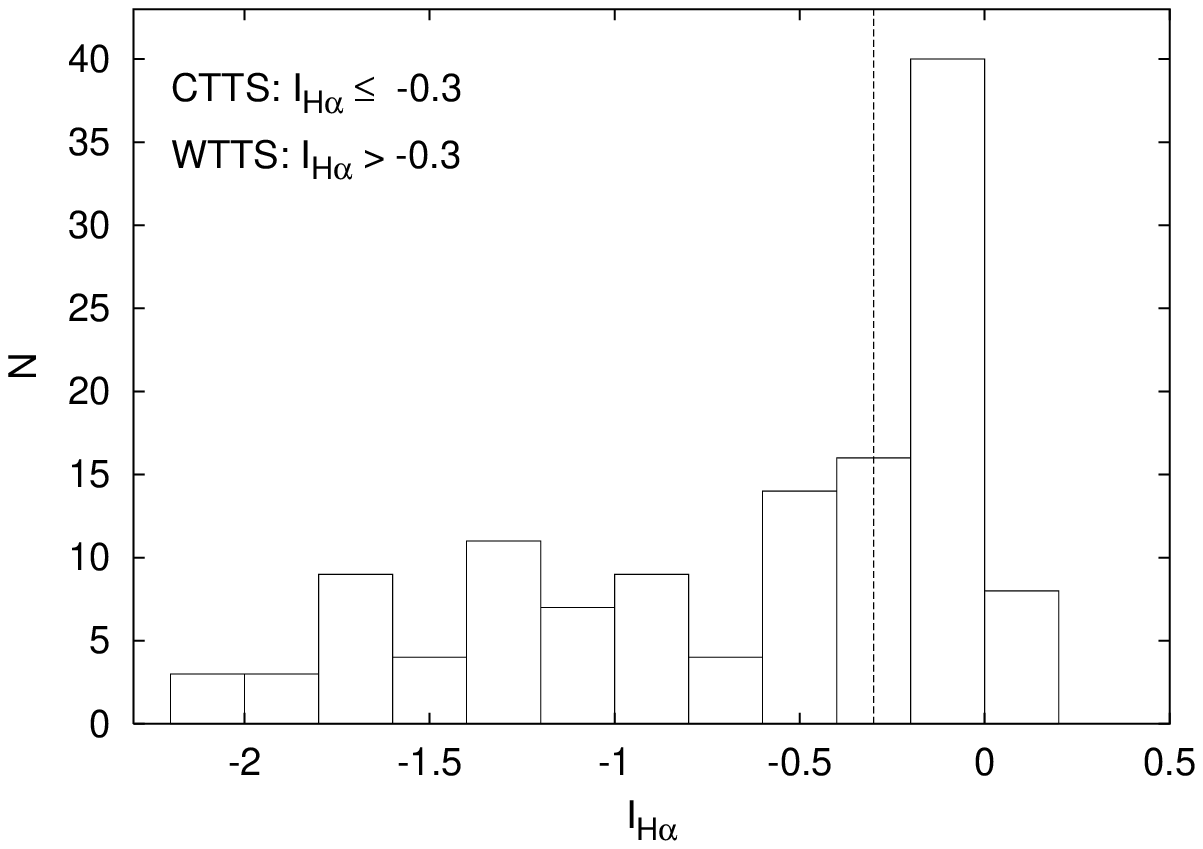}{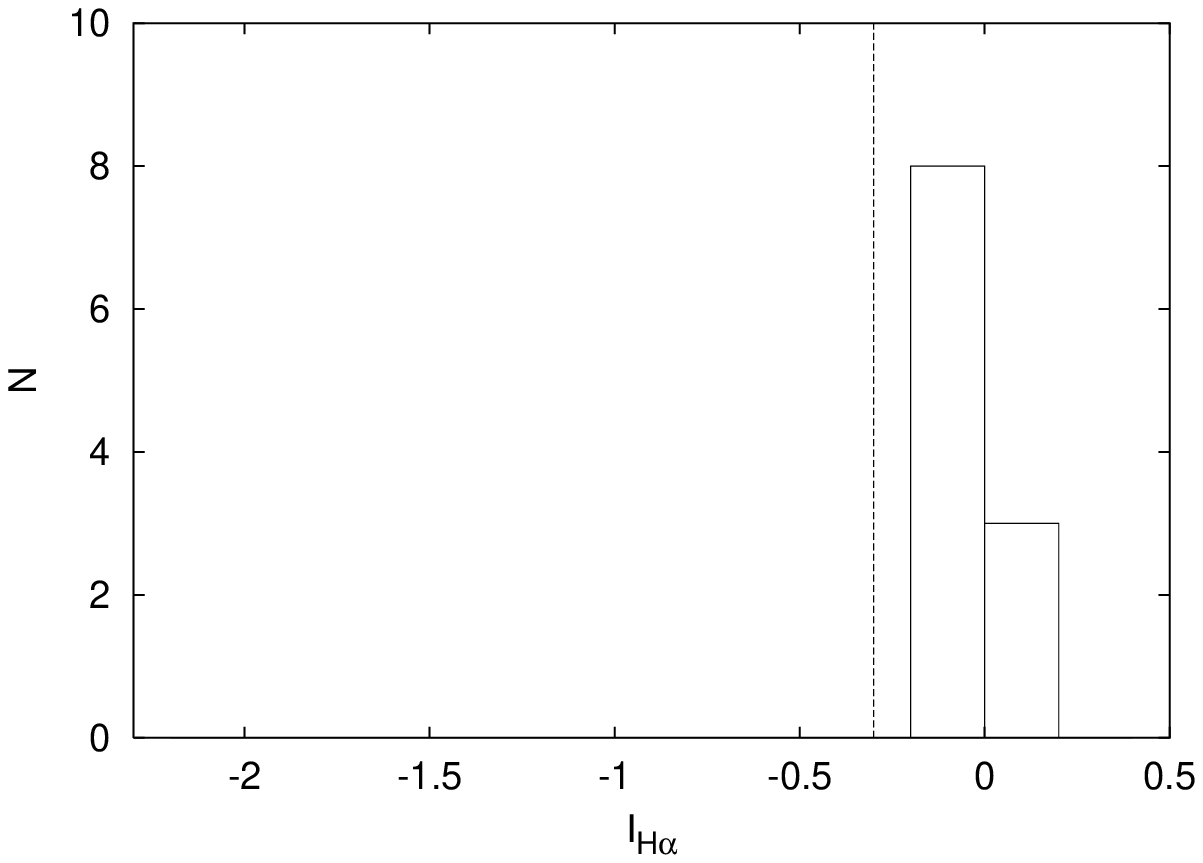}
\caption{Histogram of the $H\alpha$ indices of late type PMS stars in the Taurus -Auriga molecular cloud (left panel) and our late type PMS candidates (right panel). The vertical dashed line denotes the boundary between the cTTs and wTTs stars}
\label{fig:4}
\end{center}
\end{figure}

\citet{walt87} and \citet{walt88} divide the low mass PMS stars into two groups according to their spectroscopic and photometric properties. Classical T Tauri stars (cTTs) have strong $H\alpha$ emission, with $EW(H\alpha) > 10$\AA; wTTs have $EW(H\alpha) < 10$\AA. Most cTTs have other strong emission lines from HI, CaII and sometimes HeI. Many have [NII], [SII] and [OI] emission from winds or collimated jets. cTTs also display large infrared excesses from a circumstellar disk. Optical veiling is observable in the spectra of cTTs which indicates the presence of extra emission from a boundary layer or accretion hot spot \citep{bert89}. wTTs often have CaII and other chromospheric emission lines. They usually have nearly blackbody spectra and rarely display jet or wind emission features \citep{bert89}. 

To place our emission line stars in context with other PMS stars, we compare our sample with stars in the well studied Taurus-Auriga cloud \citep[e.g.,][]{keny95}. Fig 4. compares the distribution of the $H\alpha$ indices of T Tauri stars in Taurus-Auriga  \citep{keny98} with late type PMS stars in the field of NGC~6871. The two histograms peak at the same index value. Compared to Taurus-Auriga, NGC~6871 has few strong emission stars and an overabundance of weak emission line stars.

 With age of $\simeq 10^6$ yr, the PMS stars in Taurus-Auriga are much younger than the apparent age ($\simeq 10^7$ yr) of PMS stars in NGC~6871. Observations of other young clusters suggest that the disk accretion which powers cTTs emission declines on timescales of $\simeq 10^7$ yr \citep{hart98, hais01}. Our apparent discovery of wTTs in NGC~6871 is consistent with these observations. We plan to return to this issue in future papers where we analyze the non emission line stars in the cluster.

\section{Conclusions}

We have begun a spectroscopic survey of the galactic open cluster NGC~6871. Our spectra cover $3700-7400$\AA~at $\simeq 6$\AA~resolution. The survey is complete to $V=14.9$ and includes stars with $V < 16.5$ between the ZAMS and $10^7$ yr PMS isochrone. 

Using $H\alpha$ and $H\gamma$ spectral indices, we identify 44 emission line stars. The spectral types of these stars range from O9 - K8; one star may have an M5 spectral type. Eight stars have [NII] emission; six stars have [SII] emission. A few stars may have Li I $\lambda6708$ absorption; higher resolution spectra with good signal-to-noise would test these detections.

We use the reddening $E_{B-V}$ to divide the emission line sample into foreground stars, background stars, and cluster members. Nine stars have very low reddening and are probably foreground stars. Eighteen stars are much more highly reddened than the cluster stars. However, most of these lie above the cluster main sequence on the HRD, suggesting that these stars are cluster stars with extra (circumstellar) reddening. We thus conclude that 24 emission line stars are cluster members.

The 9 foreground stars fall into two groups. Four stars have essentially no reddening and are therefore nearby stars with $d \lesssim 300$ pc. These might be part of a nearby association, similar to the TW Hydrae association \citep{reza89,greg92,zuck01} or the $\eta$ Chamaeleontis cluster  \citep{mam99}. The other five stars also appear to lie at the same distance and might form another association or group along the line of sight to NGC~6871.

The emission line stars in the cluster consist of two groups. Stars with spectral type $\lesssim$ A0 lie on the main sequence and have ages of $\rm t < 1.4x10^7$ yr \citep{siess00}. Some of these stars may be classical Be stars. Near infrared data would allow a test of the evolutionary status of these stars. \citet{lada92} found that PMS Be stars and classical Be stars occupy well-defined, different regions of the J-H, H-K color-color diagram. A few F-type emission line stars may belong to this group. Several stars with spectral types A - G lie above the cluster main sequence and may be PMS stars. The distribution of $H\alpha$ indices for these stars coincides with the peak observed for wTTs in the Taurus-Auriga molecular cloud. If these are PMS stars, they are wTTs not cTTs. Some of these stars show [NII], [SII] emission and may have LiI absorption, which provides some evidence for their PMS status.

The discovery of [SII] and [NII] in apparent wTTs in NGC~6871 is a new and unexpected result of our survey. Previous observations of wTTs in other regions rarely detect [NII] or [SII] emission \citep{keny98,harti95}. We detect [NII] emission in $45\%$ and [SII] emission in $36\%$ of our candidate wTTs; none of these show [OI] emission features in their spectra. In many PMS stars, [OI], [NII], and [SII] emission lines are associated with disk accretion phenomena \citep[e.g. jets and disk-wind interactions;][]{hart89,harti95,rei01}. As a class, wTTs show very little evidence for a wind; the lack of an infrared excess indicates that they have already lost their disk. Thus the origin of [NII] and [SII] emission in our candidate wTTs is uncertain: If this emission is associated with these stars, the emission must be excited by some mechanism other than accretion. If the emission is not associated with these stars, a faint HII region in NGC 6871 is a plausible alternative source of forbidden line emission. Our test for nebular background emission in Section 4 appears to rule out this source of emission, but high resolution spectra would provide a better test.

\acknowledgments
We thank the Fred Lawrence Whipple Observatory staff and A. Mahdavi, K, Rines, E. Falco, W. Brown for helping with the observations and Susan Tokarz and the SAO Telescope Data Center for assistance in the data reduction. Comments from an anonymous referee improved our presentation.

\clearpage
\begin{deluxetable}{lcccc}
\tablecolumns{5} 
\tablewidth{0pc} 
\tablecaption{Absorption indices used for measuring spectral types} 
\tablehead{ 
\colhead{Index name} & \colhead{central wavelength} & \colhead{bandwidth} & \colhead{blue sideband} & \colhead{red sideband}
}
\startdata 
CN1& 4160.875& 35.00& 4100.125& 4265.375\\
CN2& 4160.875& 35.00& 4091.375& 4265.375\\
Ca$\lambda$4227&  4229.750& 12.50& 4216.625& 4247.250\\
G$\lambda$4300&   4300.125& 35.00& 4275.750& 4328.250\\
Fe$\lambda$4383&  4396.000& 51.25& 4366.000& 4450.375\\
Ca$\lambda$4455&  4464.625& 22.50& 4451.500& 4485.875\\
Fe$\lambda$4531&  4538.000& 45.00& 4510.500& 4571.125\\
Fe$\lambda$4678&  4678.375& 86.25& 4622.125& 4750.875\\
H$\beta$&   4861.250& 28.75& 4837.875& 4884.125\\
Fe$\lambda$5015&  5015.875& 76.25& 4962.125& 5059.625\\
Mg1&     5101.625& 65.00& 4926.375& 5333.625\\
Mg2&     5173.875& 39.50& 4926.375& 5333.625\\
Mgb& 5176.375& 32.50& 5152.000& 5198.875\\
Fe$\lambda$5270&  5265.650& 40.00& 5240.650& 5301.900\\
Fe$\lambda$5335&  5332.125& 40.00& 5310.250& 5358.375\\
Fe$\lambda$5406&  5401.250& 27.50& 5381.875& 5420.000\\
Fe$\lambda$5709&  5710.250& 23.75& 5686.500& 5731.500\\
Fe$\lambda$5782&  5788.375& 20.00& 5772.125& 5806.500\\
NaD& 5894.875& 32.50& 5869.875& 5936.875\\
TiO$\lambda$5968& 5968.3& 40& 5835.6& 6073.3\\
TiO$\lambda$6234& 6233.7& 40& 6106.5& 6396.8\\
H$\alpha$&  6565& 30& 6330& 6600\\
H$\beta$&   4861& 20& 4785& 5050\\
H$\gamma$&  4340& 20& 4270& 4400\\
H$\delta$&  4101& 20& 4015& 4270\\
H$\lambda$3889&   3889& 20& 3860& 3910\\
H$\lambda$3835&   3835& 20& 3815& 3860\\
He$\lambda$4144&  4144& 20& 4063 & 4175\\
He$\lambda$4387&  4387& 20& 4370& 4510\\
He$\lambda$4471&  4471& 20& 4370& 4510\\
He$\lambda$4026&  4026& 20& 3995& 4063\\
He$\lambda$4922&  4922& 20& 4900& 4950\\
CaII&    3933& 20& 3910& 4015\\
Mg$\lambda$5175&  5175& 30& 5050& 5300\\
Na$\lambda$5892&  5892& 30& 5820& 6100\\
CH$\lambda$4305&  4305& 20& 4270& 4400\\
Ba$\lambda$6495&  6495& 30& 6330& 6620\\
TiO$\lambda$6180& 6182& 40&  6127& 6372\\
TiO$\lambda$7100& 7100.5& 40& 7050& 7400\\

\enddata 
\end{deluxetable} 
\clearpage
\begin{deluxetable}{lcccccccccc}
\rotate 
\tabletypesize{\scriptsize}
\tablecolumns{11} 
\tablewidth{0pc} 
\tablecaption{Stars with certain $H\alpha$ emission in NGC~6871. The photometric data are from \citet{mass95}. } 
\tablehead{ 
\colhead{Id} & \colhead{SIMBAD Id} & \colhead{V} & \colhead{B-V} & \colhead{U-B} & \colhead{Sp. Type (present paper)} &\colhead{Sp. Type (literature)}& $\rm I_{H\alpha}$&  \colhead{E(B-V)} &\colhead{evol. stat.}&\colhead{first reported}
}
\startdata 
0001 & HD 190918 & 6.83 & 0.11 & -0.74 & - & WN5 + O9.5 III &-0.282791& $>0.41$& super-giant &  \citet{all43}\\
0006 & HD 227611 & 8.74 & 0.31 & -0.67 & B0 & B0 II:pe &-1.411940& 0.61 &giant& \citet{hilt56}\\
0008 & SAO 69404 & 8.85 & 0.17 & -0.53 & B0 &B0.5 Ve & -0.004206&0.47 &MS& \citet{grig88}\\ 
0186 & Cl* NGC 6871 BP 2  & 11.98 & 0.32 & -0.13 &B4& B5 Ve & -0.357927 &0.48&MS& \citet{bern01} \\
2352 & Cl* NGC 6871 BP 3 & 12.75 & 0.38 & -0.09 & B5 & B8 Ve &-0.333665&0.52&MS& \citet{bern01} \\
2905 & HD 226868 & 8.81 & 0.83 & -0.29 & O9& O9.7 Iab&-0.042580&$>1.13$&super-giant& \citet{mass95} \\
0215 & [MJD95] J200700.98+354006.3 & 10.27 & 0.25 & -0.51 & B1& &-0.168169&0.51&MS& present paper \\
2209 & [MJD95] J200512.93+355113.3 & 13.13 & 0.80 & 0.14 & B5&  &0.003828&0.94&PMS& present paper \\
1062 & [MJD95] J200728.78+354632.8 & 11.71 & 0.47 & -0.58 & B6&  &-0.507890&0.6&PMS& \citet{mass95} \\
1238 & [MJD95] J200711.83+353853.3 & 13.47 & 0.52 & 0.16 & B6&  &-0.023224&0.65&MS& present paper \\
0038 & [MJD95] J200547.79+353811.4 & 11.91 & 0.31 & 0.06 & B8&  &0.098522&0.4&PMS& present paper \\
2320 & [MJD95] J200453.31+353646.8 & 14.70 & 0.71 & 0.42 & A0& &-0.183464&0.71&PMS& present paper \\
2492 & [MJD95] J200418.58+355426.0 & 13.02 & 0.68 & 0.45 & A0&  &-0.038215&0.68&PMS& present paper \\
2139 & [MJD95] J200522.04+353719.8 & 13.83 & 0.46 & 0.20 & A1&  &0.251928&0.43&PMS& present paper \\
2399 & HD 227480 & 9.26 & 0.09 & -0.04 & A2&  A3 II &0.160104&0.03&MS& present paper \\
1694 & [MJD95] J200615.06+354101.4 & 14.41 & 0.76 & 0.36 & A9&  &0.177453&0.49&PMS& present paper \\
2646 & [MJD95] J195957.80+351908.6 & 15.88 & 0.96 & 0.47 & F7&  &-0.104228&0.46&PMS& present paper \\
2631 & [MJD95] J200002.77+351856.2 & 16.03 & 0.94 &-0.11 &  F8:&  &-0.109664&0.42&PMS& present paper \\
2055 & [MJD95] J200532.16+353809.0 & 15.95 & 0.85 & 0.35 & F8:&  &-0.604574&0.33&MS& present paper \\
1755 & [MJD95] J200608.03+354011.6 & 15.31 & 0.73 & 0.17 & F8&  &0.082073&0.21&PMS& present paper \\
1552 & [MJD95] J200631.73+353609.9 & 15.37 & 0.77 & 0.32 & F9&  &-0.077949&0.22&MS& present paper \\
2675 & [MJD95] J195948.26+352142.9 & 15.92 & 0.98 & 0.38 & G0:&  &0.035220&0.4&PMS& present paper \\
2245 & [MJD95] J200508.36+353708.5 & 13.98 & 1.15 & 0.25 & G1&  &-0.028562&0.55&PMS& present paper \\
2621 & [MJD95] J200006.39+351402.8 & 14.79 & 1.15 & 0.82 & G5&  &-0.020041&0.49&PMS& present paper \\
2138 & [MJD95] J200522.08+354212.0 & 14.40 & 1.21 & 0.87 & G9&  &-0.019457&0.43&PMS& present paper \\ 
1286 & [MJD95] J200705.96+355110.7 & 14.08 & 1.48 & 1.29 & K2&  &0.000891&0.59&giant& present paper \\
1647 & [MJD95] J200621.20+355600.6 & 14.42 & 1.41 & 1.05 & K8&  &-0.081935&0.02&PMS& present paper \\
1659 & [MJD95] J200619.32+355546.4 & 14.37 & 1.38 & 0.86 & K8&  &-0.073231&-0.01&PMS& present paper \\
\enddata 
\end{deluxetable} 

\begin{deluxetable}{lcccccccccc} 
\rotate
\tabletypesize{\scriptsize}
\tablecolumns{11} 
\tablewidth{0pc} 
\tablecaption{Emission line candidate stars in NGC~6871. The photometric data are from \citet{mass95}} 
\tablehead{ 
\colhead{Id} & \colhead{MJD95 Id} & \colhead{V} & \colhead{B-V} & \colhead{U-B} &\colhead{Sp. Type (present paper)} &\colhead{Sp. Type (literature)}&  $\rm I_{H\alpha}$& \colhead{E(B-V)} &\colhead{evol. stat.}& \colhead{first reported}
}
\startdata 
2 & HD 190919 & 7.29 & 0.25 & -0.62 & O9& B1 Ib &0.085103&$>0.55$&super-giant& present paper \\
3 & SAO 69403 & 7.38 & 0.25 & -0.64 & B0& B1 Ib &0.082967&0.55&super-giant& present paper \\
2092 & [MJD95] J200557.85+354721.1 & 11.93 & 0.23 & -0.07 & B4&  &0.154545&0.39& MS& present paper \\
1651 & [MJD95] J200622.09+354847.8 & 13.90 & 0.52 & 0.34 & A0&  &0.240811&0.52& PMS& present paper \\
1980 & [MJD95] J200544.49+355352.3 & 14.34 & 0.48 & 0.23 & A2&  &0.277399&0.42& MS& present paper \\
1093 & [MJD95] J200725.36+353633.7 & 16.15 & 0.91 & 0.29 & A2&  &0.287980&0.85& MS& present paper \\
2637 & [MJD95] J200000.73+352316.3 & 15.02 & 0.65 & 0.46 & A3&  &0.298494&0.56& MS& present paper \\
1710 & [MJD95] J200613.34+355244.8 & 15.24 & 0.59 & 0.31 & A5:&  &0.245813&0.45& PMS& present paper \\
2156 & [MJD95] J200519.35+354222.6 & 14.35 & 0.72 & 0.44 & A7&  &0.210561&0.53& PMS& present paper \\
2067 & [MJD95] J200529.55+354551.7 & 14.31 & 0.74 & 0.55 & F1&  &0.165404&0.41& PMS& present paper \\
1379 & [MJD95] J200656.29+354947.8 & 16.47 & 1.19 & -0.85 & F2&  &0.129417&0.84& PMS& present paper \\
1845 & [MJD95] J200559.51+355137.2 & 14.81 & 1.01 & 0.34 & F6&  &0.099795&0.55& PMS& present paper \\
1221 & [MJD95] J200714.35+355322.2 & 15.18 & 0.65 & 0.21 & F7&  &0.090169&0.15& MS& present paper \\
2545 & [MJD95] J200408.60+355006.3 & 15.63 & 0.76 & 0.06 & F8&  &0.102688&0.24& MS& present paper \\
1869 & [MJD95] J200556.66+354229.7 & 15.75 & 0.82 & 0.10 & F9&  &0.076941&0.27& MS& present paper \\
113 & [MJD95] J200608.60+353458.1 & 13.02 & 1.83 & 1.97 & M5:&  &-0.022871&0.01& PMS& present paper \\
\enddata 
\end{deluxetable} 

\end{document}